\documentclass[showpacs,preprintnumbers,amsmath,amssymb,preprint]{revtex4}
\usepackage{graphicx}
\usepackage{dcolumn}
\usepackage{bm}

\newcommand{\GeV}{\ensuremath{\mathrm{Ge\kern -0.1em V}}}
\newcommand{\MeV}{\ensuremath{\mathrm{Me\kern -0.1em V}}}

\begin{document}

\begin{flushright}
submitted to PRD
\end{flushright}
\vspace*{0.50in}

\begin{center}
{\bf{The Underlying Event in Hard Interactions at the Tevatron $\bar p p$ Collider}
}
\end{center}

\font\eightit=cmti8
\def\r#1{\ignorespaces $^{#1}$}
\hfilneg
\begin{sloppypar}
\noindent
D.~Acosta,\r {14} T.~Affolder,\r 7 M.G.~Albrow,\r {13} D.~Ambrose,\r {36}   
D.~Amidei,\r {27} K.~Anikeev,\r {26} J.~Antos,\r 1 
G.~Apollinari,\r {13} T.~Arisawa,\r {50} A.~Artikov,\r {11}
% ...
\end{sloppypar}
\vskip .026in
\begin{center}
(CDF Collaboration)
\end{center}

\vskip .026in
\begin{center}
\r 1  {\eightit Institute of Physics, Academia Sinica, Taipei, Taiwan 11529, 
Republic of China} \\
\r 2  {\eightit Argonne National Laboratory, Argonne, Illinois 60439} \\
\r 3  {\eightit Istituto Nazionale di Fisica Nucleare, University of Bologna,
I-40127 Bologna, Italy} \\
\r 4  {\eightit Brandeis University, Waltham, Massachusetts 02254} \\
\r 5  {\eightit University of California at Davis, Davis, California  95616} \\
\r 6  {\eightit University of California at Los Angeles, Los 
Angeles, California  90024} \\ 
\r 7  {\eightit University of California at Santa Barbara, Santa Barbara, California 
93106} \\ 
\r 8 {\eightit Instituto de Fisica de Cantabria, CSIC-University of Cantabria, 
39005 Santander, Spain} \\
\r 9  {\eightit Carnegie Mellon University, Pittsburgh, Pennsylvania  15213} \\
\r {10} {\eightit Enrico Fermi Institute, University of Chicago, Chicago, 
Illinois 60637} \\
\r {11}  {\eightit Joint Institute for Nuclear Research, RU-141980 Dubna, Russia}
\\
\r {12} {\eightit Duke University, Durham, North Carolina  27708} \\
\r {13} {\eightit Fermi National Accelerator Laboratory, Batavia, Illinois 
60510} \\
\r {14} {\eightit University of Florida, Gainesville, Florida  32611} \\
\r {15} {\eightit Laboratori Nazionali di Frascati, Istituto Nazionale di Fisica
               Nucleare, I-00044 Frascati, Italy} \\
\r {16} {\eightit University of Geneva, CH-1211 Geneva 4, Switzerland} \\
\r {17} {\eightit Glasgow University, Glasgow G12 8QQ, United Kingdom}\\
\r {18} {\eightit Harvard University, Cambridge, Massachusetts 02138} \\
\r {19} {\eightit Hiroshima University, Higashi-Hiroshima 724, Japan} \\
\r {20} {\eightit University of Illinois, Urbana, Illinois 61801} \\
\r {21} {\eightit The Johns Hopkins University, Baltimore, Maryland 21218} \\
\r {22} {\eightit Institut f\"{u}r Experimentelle Kernphysik, 
Universit\"{a}t Karlsruhe, 76128 Karlsruhe, Germany} \\
\r {23} {\eightit Center for High Energy Physics: Kyungpook National
University, Taegu 702-701; Seoul National University, Seoul 151-742; and
SungKyunKwan University, Suwon 440-746; Korea} \\
\r {24} {\eightit Ernest Orlando Lawrence Berkeley National Laboratory, 
Berkeley, California 94720} \\
\r {25} {\eightit University College London, London WC1E 6BT, United Kingdom} \\
\r {26} {\eightit Massachusetts Institute of Technology, Cambridge,
Massachusetts  02139} \\   
\r {27} {\eightit University of Michigan, Ann Arbor, Michigan 48109} \\
\r {28} {\eightit Michigan State University, East Lansing, Michigan  48824} \\
\r {29} {\eightit Institution for Theoretical and Experimental Physics, ITEP,
Moscow 117259, Russia} \\
\r {30} {\eightit University of New Mexico, Albuquerque, New Mexico 87131} \\
\r {31} {\eightit Northwestern University, Evanston, Illinois  60208} \\
\r {32} {\eightit The Ohio State University, Columbus, Ohio  43210} \\
\r {33} {\eightit Osaka City University, Osaka 588, Japan} \\
\r {34} {\eightit University of Oxford, Oxford OX1 3RH, United Kingdom} \\
\r {35} {\eightit Universita di Padova, Istituto Nazionale di Fisica 
          Nucleare, Sezione di Padova, I-35131 Padova, Italy} \\
\r {36} {\eightit University of Pennsylvania, Philadelphia, 
        Pennsylvania 19104} \\   
\r {37} {\eightit Istituto Nazionale di Fisica Nucleare, University and Scuola
               Normale Superiore of Pisa, I-56100 Pisa, Italy} \\
\r {38} {\eightit University of Pittsburgh, Pittsburgh, Pennsylvania 15260} \\
\r {39} {\eightit Purdue University, West Lafayette, Indiana 47907} \\
\r {40} {\eightit University of Rochester, Rochester, New York 14627} \\
\r {41} {\eightit Rockefeller University, New York, New York 10021} \\
\r {42} {\eightit Instituto Nazionale de Fisica Nucleare, Sezione di Roma,
University di Roma I, ``La Sapienza," I-00185 Roma, Italy}\\
\r {43} {\eightit Rutgers University, Piscataway, New Jersey 08855} \\
\r {44} {\eightit Texas A\&M University, College Station, Texas 77843} \\
\r {45} {\eightit Texas Tech University, Lubbock, Texas 79409} \\
\r {46} {\eightit Institute of Particle Physics, University of Toronto, Toronto
M5S 1A7, Canada} \\
\r {47} {\eightit Istituto Nazionale di Fisica Nucleare, University of Trieste/\
Udine, Italy} \\
\r {48} {\eightit University of Tsukuba, Tsukuba, Ibaraki 305, Japan} \\
\r {49} {\eightit Tufts University, Medford, Massachusetts 02155} \\
\r {50} {\eightit Waseda University, Tokyo 169, Japan} \\
\r {51} {\eightit University of Wisconsin, Madison, Wisconsin 53706} \\
\r {52} {\eightit Yale University, New Haven, Connecticut 06520} \\
\end{center}


\date{\today}

\begin{abstract}
For comparison of inclusive jet cross sections measured at hadron-hadron colliders to next-to-leading order (NLO) parton-level calculations, the energy deposited in the jet cone by spectator parton interactions must first be subtracted. The assumption made at the Tevatron is that the spectator parton interaction energy is similar to the ambient level measured in minimum bias events.
In this paper, we test this assumption by measuring the ambient charged track momentum in events containing large transverse energy 
jets at $\sqrt{s}=1800$ GeV and $\sqrt{s}=630$ GeV and comparing this ambient momentum 
with that observed both in minimum bias events and with that predicted by two 
Monte Carlo  models. Two cones in $\eta$--$\phi$ space are defined, at the same pseudo-rapidity, 
$\eta$, as the jet with the highest transverse energy ($E_T^{(1)}$), and at $\pm 90^o$ in the
azimuthal direction, $\phi$. The total charged track momentum inside each of the two cones is measured. The minimum  momentum in the two cones is almost independent of 
$E_T^{(1)}$ and is similar to the momentum observed in minimum bias events,   
whereas the  maximum momentum increases roughly linearly with the  jet $E_T^{(1)}$ over most of the measured range. 
This study will help improve the precision of comparisons of jet cross section data and NLO perturbative QCD predictions. 
The distribution of the sum of the track momenta in the two cones is also examined for five different $E_T^{(1)}$ bins. 
The HERWIG and PYTHIA Monte Carlos are reasonably successful in describing the data, 
but neither can describe completely all of the event properties. 
\end{abstract}

\maketitle

Jet production at hadron colliders, the highest energy probe
in particle physics, has been used to measure parton distribution functions, the running of the strong coupling constant, $\alpha_s$, and to 
search for new physics.  
At the Fermilab
$\bar p p$ collider, the jet production rate has been measured for jets of 15-450 GeV 
at $\sqrt{s}=1800$ GeV~\cite{CDF1},~\cite{CDF2},~\cite{CDF3},~\cite{DZero1} and jets of 
15-150 GeV at $\sqrt{s}=630$ GeV~\cite{CDF4},~\cite{DZero2}. The production of jets involves the
interaction of an individual parton (quark or gluon) from one beam hadron
with a parton from the other beam hadron. Each of the interacting partons
carries only a fraction of the parent hadron's momentum with the
residual momentum remaining with the other (spectator) constituents of
the hadron. In addition, there are interactions between the spectator
constituents of the two hadrons which normally occur at low momentum transfers. 
Measurements involving the observed
jets are compared to perturbative QCD predictions.
For NLO perturbative QCD predictions, only the parton level cross section, i.e. the
cross section of two partons producing either two or three partons in the final state, is
calculated. After convolution with the parton distribution functions, this cross section
is directly compared with experimental data.
For these comparisons to be valid,
the energy from spectator interactions, which may fall in the jet cone, must be subtracted
from the experimentally observed jets. 
In hard interaction jet events, the energy outside the two primary jets consists of energy from 
spectator
interactions (soft and semi-hard), initial and final state radiation and any hadronization
leakage from the jet cones. Initial and final state  radiation effects  are part of higher order perturbative QCD 
calculations and at least a portion of  these effects are already included in NLO calculations. 

Because the CDF detector measures the momenta of low $P_T$ tracks more accurately than the calorimeter measures their  energies, we choose to work with the track momenta in our analysis. 
 We will call the charged track momenta associated with spectator interactions the 
{\it underlying event momentum}. 
It is the momentum
in an event which is not directly related to the hard interaction. Clearly, this is a working
definition as a coupling exists between all aspects of a $\bar p p$ interaction. 
For example, the hadronization of the partons from the hard interaction and from the 
spectator interactions are ultimately linked as the final state hadrons must be colorless.
In current QCD studies at hadron colliders, the underlying event energy in the jet events is assumed to be
 well approximated by the ambient energy in the events collected with minimal trigger requirements.
Normally, these {\em minimum bias} events are triggered by presence of particles away from the beam
in the forward and  backward direction.
The subtraction of the underlying event energy leads to the largest uncertainty in
jet cross section measurement for $E_T\le 50$ GeV~\cite{CDF1}. A precise measurement of the 
spectator interaction energy is essential for the modeling/understanding of non-perturbative 
QCD effects and for any quantitative improvement of the jet studies. Another important question is 
whether the presence
of a hard interaction in the event influences the spectator interactions.

The measurement of the momentum in minimum bias events is important in its own right as it is used
to estimate the effect
of pile-up events on any signal at hadron colliders, where,  due to high instantaneous
luminosity, several interactions may occur in the same bunch crossing.
In this paper, we  present a measurement of the momentum deposited far 
from the jets in $\bar p p$ interactions at $\sqrt{s}=1800$ and $\sqrt{s}=630$ GeV
and compare our measurement  with the momentum observed in minimum bias events and with the predictions from two Monte Carlo models. The jet samples used in this analysis are the same as for the inclusive jet cross section measurements at the two center-of-mass energies. The study of {\it interjet} soft gluon radiation is  also of special interest in QCD as its emission originates from  the flow of color between jets. The analysis of such observables may lead  to a better understanding of color neutralization~\cite{pino,salam}.

The study reported in this paper is complementary to our previous  analysis ~\cite{Rick},
which examined the evolution of event structure in low to moderate $E_T$ events in $\bar p p$ interactions at $\sqrt{s}=1800$ GeV 
by studying charged particle jets from 0.5 GeV/c to 50 GeV/c.
The previous study found that the momentum transverse to the leading jet rises rapidly
in the 
0.5-5.0 GeV/c range and is almost constant when the leading charged particle jet has transverse momentum greater than about 10 GeV/c. 

\begin{figure}[h]
\centerline{\includegraphics[height=7.cm]{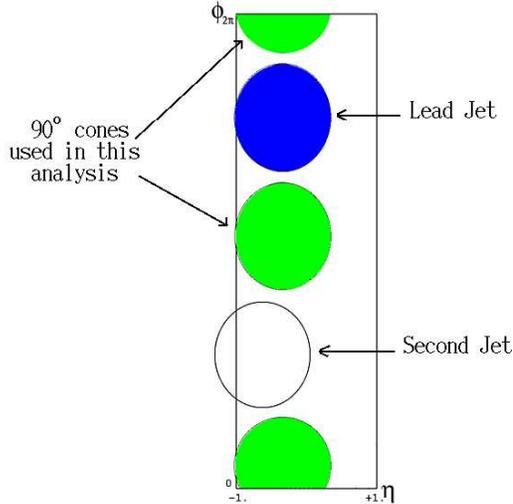}}
\caption{An example of a two jet event  in the detector region under study. The cones used for the determination of the underlying event contribution are at $\eta = \eta^{(1)}$ and $\phi = \phi^{(1)} \pm 90^{\circ}$ where
($\eta^{(1)},\phi^{(1)}$) is the centroid of the highest $E_T$ jet in the event.}
\label{Fig-phase-space}
\end{figure}

To study the underlying momentum in jet events, we define two cones with radius 
$R=\sqrt{(\Delta\eta)^2+(\Delta\phi)^2}=0.7$ centered at $\eta\!=\!\eta^{(1)}$, and
$\phi\!=\!\phi^{(1)}\pm 90 ^0$ where ($\eta^{(1)}$, $\phi^{(1)}$) is the centroid of the 
highest energy jet in the event as shown in Figure~\ref{Fig-phase-space}. The sum of the transverse momenta of all tracks in the two cones
is labeled $P_T^{90,min}$ and  $P_T^{90,max}$, where  $P_T^{90,max}$ is higher of the two values. 
By definition, $P_T^{90,max}$ should contain a larger  
contribution from initial and final state radiation than  $P_T^{90,min}$. In the approximation of a
negligible four parton final state component, $P_T^{90,min}$ is a measure of the underlying momentum 
in the jet event. In minimum bias data, we perform a similar analysis but with the cone centroid selected randomly in the central rapidity region, $|\eta|<0.5$. 
We also use a second procedure, the {\it Swiss cheese} method, in which the transverse momenta of all the  tracks except those in the two or three highest energy jets are summed and compared with Monte Carlo predictions and minimum bias data. A study of this type was first suggested in Ref.~\cite{wm}.
 

The data were collected using the CDF detector~\cite{CDF-detector} with the Fermilab Tevatron Collider at
$\sqrt{s}=1800$ GeV (1994-1995) and $\sqrt{s}=630$ GeV (1995). The CDF detector is a multipurpose detector
consisting of a tracking system in a solenoidal magnetic field, calorimeters, muon chambers
and two arrays of scintillator counters (BBC) located at $\pm 5.8 $m from the nominal interaction point along
the beam direction, covering the $3.2\le |\eta|\le 5.9$ region.
Minimum bias events were triggered by a coincidence of hits in these counters.
The BBC cross section is 51.15$\pm 1.60$  mb compared to a total inelastic cross section for $\bar p p$ interactions of $60.33 \pm 1.40$ mb at $\sqrt{s}=1800$ GeV~\cite{CDF_tot}.
The jet data were collected using four triggers requiring a cluster of energy in the calorimeter with 
$E_T\ge$ 20, 50, 70 and 100 GeV at $\sqrt{s}=1800$ GeV. These data samples, and the  jet clustering and energy
corrections have been described in detail in ~\cite{CDF1}. The data at $\sqrt{s}=630$ GeV employed two
triggers requiring a cluster with $E_T\ge$ 5 and 15 GeV respectively. The jet energies were corrected for
any energy loss in the detector. At both energies we use only those events in which the centroid of the highest energy jet
is within the central rapidity region, $|\eta|<0.5$. 

The tracking system consists of a silicon vertex detector (SVX$^{'}$), 
a vertex tracking chamber (VTX), and a central tracking chamber (CTC)
~\cite{CDF_tracking}. 
The vertex reconstruction is performed using information from the VTX and the CTC.
In this analysis, the jet events  were required to have one and only one primary vertex of
high quality (corresponding to a high track multiplicity). For inclusive jet analyses in general, there is no restriction on the number of vertices as long as there is at least one high quality vertex. The one vertex requirement in this analysis is implemented in order to restrict the events to
those  in which only one interaction occurred  during that beam crossing. 
  For minimum bias data, the requirement is changed to
one vertex of medium quality (corresponding to a lower minimum track multiplicity, but one resulting from beam-beam rather than beam-gas interactions). 

Track reconstruction takes place primarily using hit information from the CTC. In order to
ensure a high quality for the reconstructed tracks, each track is required to
have at least four  hits in each of the five axial super-layers and hits in at
least one stereo super-layer.
The momentum resolution in the rapidity region $|\eta| \le 1$ is better than 
$\delta P_T/P_T{^2}\leq 0.002$ (GeV/c)$^{-1}$.
We require the tracks to have $P_T\ge 0.4$ GeV/c and to be within 5 cm in the longitudinal and 0.5 cm 
in the transverse
direction of the $\bar p p$ vertex. The uncertainty on the quantities measured in this analysis is evaluated by 
loosening these cuts to 10 cm and 5 cm respectively.
The charged track reconstruction efficiency is uniform in  rapidity
for  $|\eta|\le 1$ and is on average $\sim 92\pm3\%$~\cite{warburton}.
The efficiency  drops to 80\% at low $P_T$  (0.4-0.5 GeV/c) and to 60\% in the region
$1.0\le|\eta|\le 1.2$.
We correct the  data for the inefficiency in both regions. 
The main systematic uncertainty in our analysis arises from the track selection
criteria and from the track reconstruction efficiency.
The data were compared to Monte Carlo predictions from the
programs HERWIG~\cite{HERWIG} and PYTHIA~\cite{PYTHIA}. 
At $\sqrt{s}=1800$ GeV, four samples of jet events were generated with HERWIG and PYTHIA, with a
minimum transverse momentum for the hard scattering of 20, 40, 60 and 80 GeV, for the four samples.
The leading jet in the generated distributions was required to have a transverse
energy of 40, 75, 100 and 130 GeV, respectively. 
The output from both Monte Carlo programs consists of the 4-vectors of the final state hadrons. 
For comparison to the Monte Carlo predictions, the data were corrected for the track reconstruction efficiency.

\begin{figure}[h]
\centering
\mbox{
\includegraphics[height=11.cm]{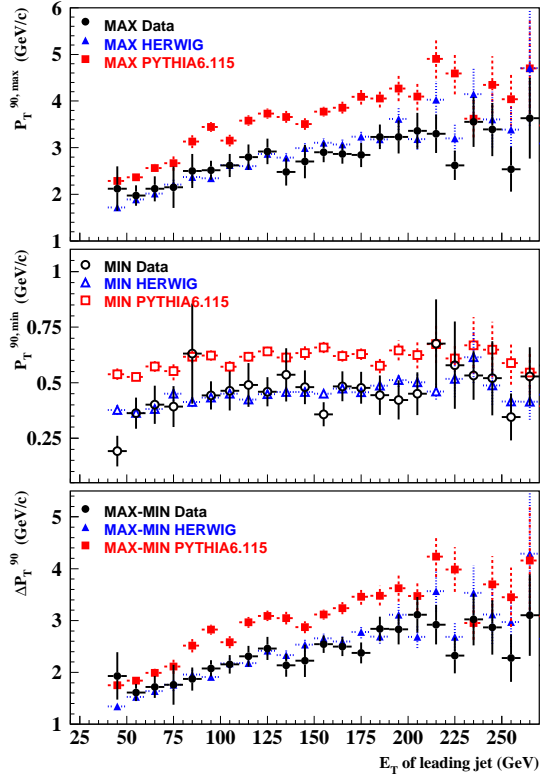}}
\caption{$P_T^{90,max}$, $P_T^{90,min}$ and their difference $\Delta P_T^{90}$ as a function of the $E_T$  of the highest energy jet at $\sqrt{s}=1800$ GeV.}
\label{Fig_track_max_min}
\end{figure}

In Figure~\ref{Fig_track_max_min}, a comparison of the $P_T^{90,max/min}$ distributions between data and Monte Carlo is shown. $P_T^{90,max}$ increases as the leading jet $E_T$ increases, both in the data and in the Monte Carlo predictions. 
The $P_T^{90,min}$ distributions are almost independent of $E_T^{(1)}$ indicating that any contribution
from higher order radiation, at least in this $\eta$--$\phi$ region, is small.  
Both the Monte Carlo predictions and the data show a similar
behavior. The average values of $P_T^{90,max}$ and  $P_T^{90,min}$  are given in 
Table~\ref{Table-max-min-ptrack} for different intervals of $E_T^{(1)}$, for the data and 
for the two Monte Carlos. Good agreement is observed with HERWIG, while  
PYTHIA lies above the data.
\begin{table}[t]
\begin{center}
\caption{Average $P_T$ inside the $max$ and $min$ cone at  $\eta = \eta^{(1)}$ and
$\phi = \phi^{(1)} \pm 90^\circ$ for  $\sqrt{s}=1800$ GeV data. 
In data, the first errors shown are statistical and the second are systematic.}
\label{Table-max-min-ptrack}
\begin{tabular}{|c|c|c|c|c|}
\hline
\multicolumn{1}{|c|}{$E_T^{(1)}$}&
\multicolumn{1}{|c|}{DATA} &
\multicolumn{1}{|c|}{HERWIG} &
\multicolumn{1}{|c|}{PYTHIA} &
\multicolumn{1}{|c|}{PYTHIA} \\
\multicolumn{1}{|c|}{(GeV)}   &
\multicolumn{1}{|c|}{ } &
\multicolumn{1}{|c|}{ } &
\multicolumn{1}{|c|}{(default)} &
\multicolumn{1}{|c|}{(tuned)} \\
\hline
\hline
\multicolumn{5}{|c|}{$P_T^{90,max}$ (GeV/c)} \\
\hline
 40-80  & 2.04$\pm$0.09$\pm$0.21 & 1.92$\pm$0.04 & 2.43$\pm$0.04 & 2.19$\pm$0.04  \\
 80-120 & 2.64$\pm$0.09$\pm$0.19 & 2.49$\pm$0.05 & 3.39$\pm$0.06 & 2.96$\pm$0.06  \\
 120-160& 2.89$\pm$0.09$\pm$0.22 & 2.95$\pm$0.05 & 3.69$\pm$0.06 & 3.56$\pm$0.06  \\
 160-200& 3.27$\pm$0.10$\pm$0.22 & 3.21$\pm$0.07 & 4.02$\pm$0.09 & 3.93$\pm$0.09  \\
 200-270& 3.64$\pm$0.21$\pm$0.24 & 3.59$\pm$0.16 & 4.35$\pm$0.17 & 4.24$\pm$0.19  \\
\hline
\hline
\multicolumn{5}{|c|}{$P_T^{90,min}$ (GeV/c)} \\
\hline
40-80     & 0.37$\pm$0.03$\pm$0.06 & 0.38$\pm$0.01 & 0.54$\pm$0.01 & 0.38$\pm$0.01 \\
80-120    & 0.47$\pm$0.02$\pm$0.07 & 0.43$\pm$0.01 & 0.61$\pm$0.01 & 0.44$\pm$0.01 \\
120-160   & 0.42$\pm$0.02$\pm$0.06 & 0.45$\pm$0.01 & 0.64$\pm$0.01 & 0.48$\pm$0.01 \\
160-200   & 0.46$\pm$0.02$\pm$0.06 & 0.48$\pm$0.01 & 0.62$\pm$0.02 & 0.53$\pm$0.02 \\
200-270   & 0.53$\pm$0.05$\pm$0.07 & 0.50$\pm$0.02 & 0.63$\pm$0.03 & 0.53$\pm$0.04 \\
\hline
\end{tabular}
\end{center}
\end{table}   
\begin{table}[t]
\begin{center}
\caption{Monte Carlo's default and tuned parameters. MSTP(82) defines the structure of the multiple parton interactions; PARP(82) is the regularization scale of the  transverse momentum spectrum for multiple interactions (with MSTP(82)$\ge$2); PARP(85) and PARP(86) are the probability that the multiple interaction produces two gluons with color connections to the nearest neighbors (or as a closed gluon loop)~\cite{PYTHIA}. }
\label{Table-MC}
\begin{tabular}{|c|c|c|c|c|}
\hline
\multicolumn{1}{|c|}{}&
\multicolumn{3}{|c|}{PYTHIA} &
\multicolumn{1}{|c|}{HERWIG} \\
\cline{2-5}
\multicolumn{1}{|c|}{} &
\multicolumn{1}{|c|}{default 6.115} &
\multicolumn{1}{|c|}{tuned 6.115}   &
\multicolumn{1}{|c|}{tuned 6.115} &
\multicolumn{1}{|c|}{default 5.6 } \\
\multicolumn{1}{|c|}{} &
\multicolumn{1}{|c|}{1800 and 630 GeV} &
\multicolumn{1}{|c|}{1800 GeV} &
\multicolumn{1}{|c|}{630 GeV} &
\multicolumn{1}{|c|}{1800 and 630 GeV} \\
\hline
\hline
Parton distribution function & MRSG & CTEQ4L & CTEQ4L & CTEQ3L \\
MSTP(82) & 1 & 3 & 3 & - \\
PARP(82) & - & 2.0 GeV/c & 1.4 GeV/c & - \\
PARP(85) & 0.33 & 1 & 1 & - \\
PARP(86) & 0.66 & 1 & 1 & - \\
\hline
\end{tabular}
\end{center}
\end{table}   
\begin{figure}[!hb]
\centering
\mbox{
\includegraphics[height=11.cm]{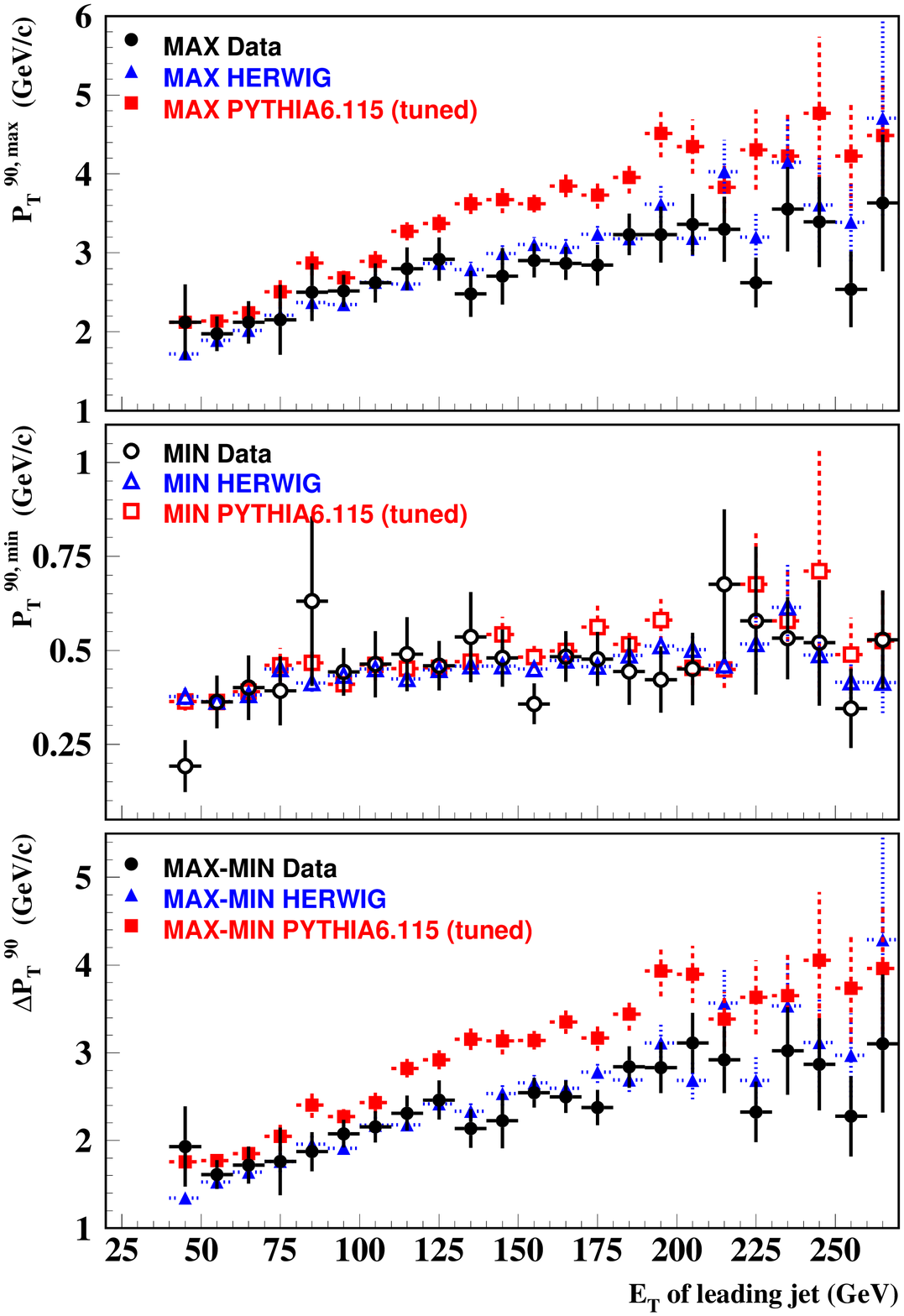}}
\caption{$P_T^{90,max}$, $P_T^{90,min}$ and their difference $\Delta P_T^{90}$ as a function of the $E_T$  of the highest energy jet at $\sqrt{s}=1800$ GeV. PYTHIA has been tuned to reproduce the data.}
\label{Fig_track_max_min_tuned}
\end{figure}
\begin{figure}[h]
\centerline{\includegraphics[height=13.cm]{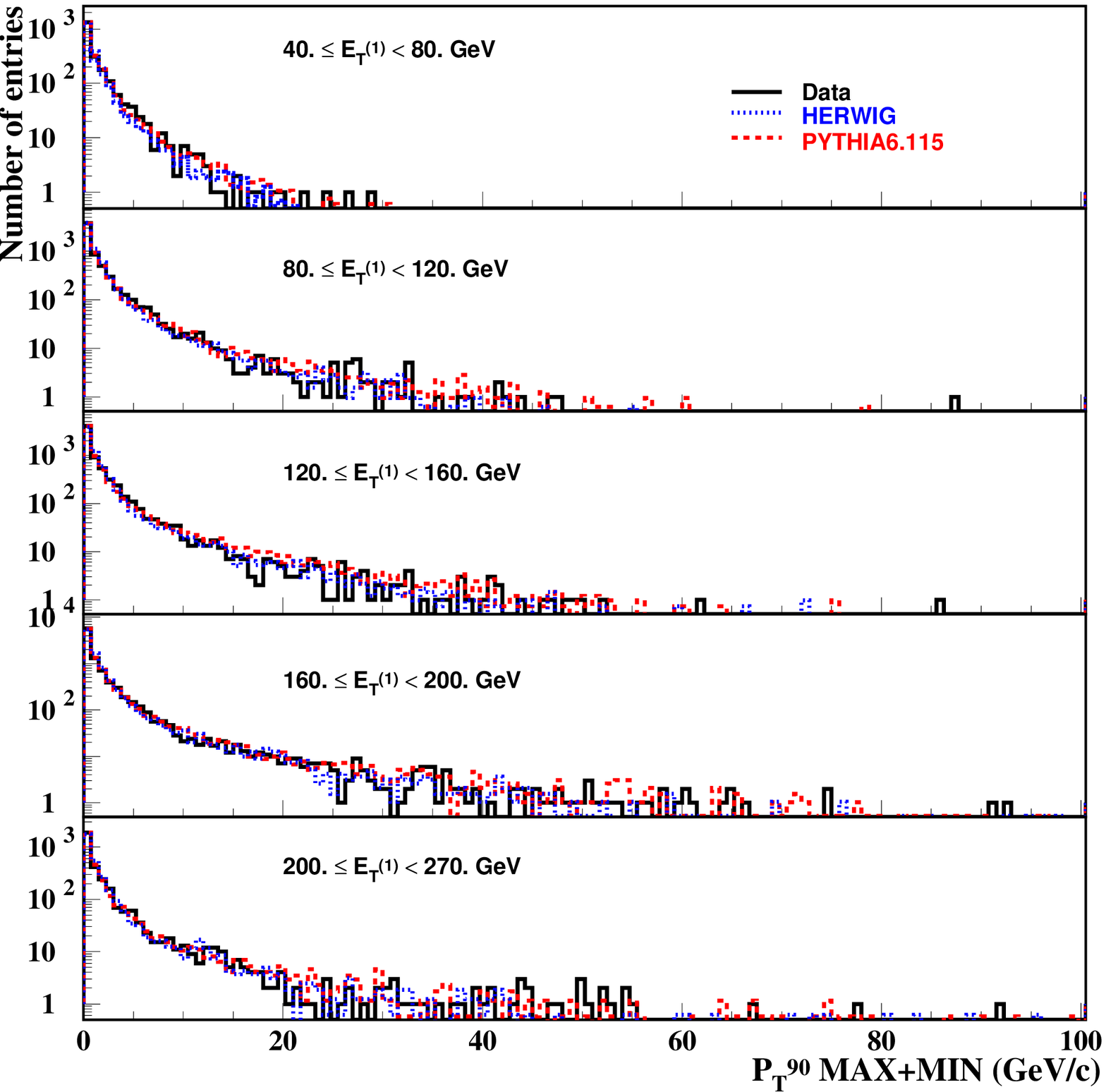}}
\caption{The distributions for the total $p_T$ in the sum of the $max$ and $min$ cones is plotted for five different bins of the $E_T$ of the highest energy jet. Data, HERWIG and PYTHIA distributions are shown at $\sqrt{s}=1800$ GeV. }
\label{Fig_max_dist}
\end{figure}
The parameters of the underlying event model in PYTHIA can be adjusted more
easily than those in HERWIG. We have attempted to reach a better agreement with
the predictions from PYTHIA by using a more modern parton distribution function 
(CTEQ4L~\cite{CTEQ4L} instead of MRSG~\cite{MRSG}), using the option of varying impact 
parameters with a matter distribution inside the hadron described by a simple Gaussian (MSTP(82)=3), 
and by decreasing the regularization scale of the transverse momentum spectrum for multiple 
interactions (P$_{T0}$) to 2.0 GeV/c from the default value of 2.3 GeV/c. (Such a decrease causes the double parton scattering component of the underlying event to be less hard, leading to a better agreement with the data.)
Table~\ref{Table-MC}  summarizes the Monte Carlo's default and tuned parameters.
The behavior of PYTHIA with the adjusted parameters can be observed in 
Figure~\ref{Fig_track_max_min_tuned} and in the last column in 
Table~\ref{Table-max-min-ptrack}. The tuning leads to a better agreement with the data in the low $E_T^{(1)}$ region but leaves PYTHIA  still somewhat larger in the high $E_T^{(1)}$ region.

In Figure~\ref{Fig_max_dist}, the total charged track momentum in the two cones ($min$+$max$) is shown for five different bins of $E_T^{(1)}$.
The effects due  to  large angle (away from any jet) soft gluon emission are expected to be appreciable when the transverse momentum in the $min$+$max$ cones ($p_T^{max}+p_T^{min}$)  is larger than a few GeV and  when the ratio  of the lead jet transverse momentum to the transverse momentum in the cones is large. Such emissions are included in an approximate way in existing parton shower Monte Carlos  and a detailed comparison may lead to improvements in their treatment~\cite{pino1}. Qualitatively,  HERWIG and PYTHIA agree with the data, although the PYTHIA prediction tends to be  slightly higher than the data for larger values of transverse momentum in the two cones~\cite{TuneA}. 

\begin{figure}[h]
\centerline{\includegraphics[height=11cm]{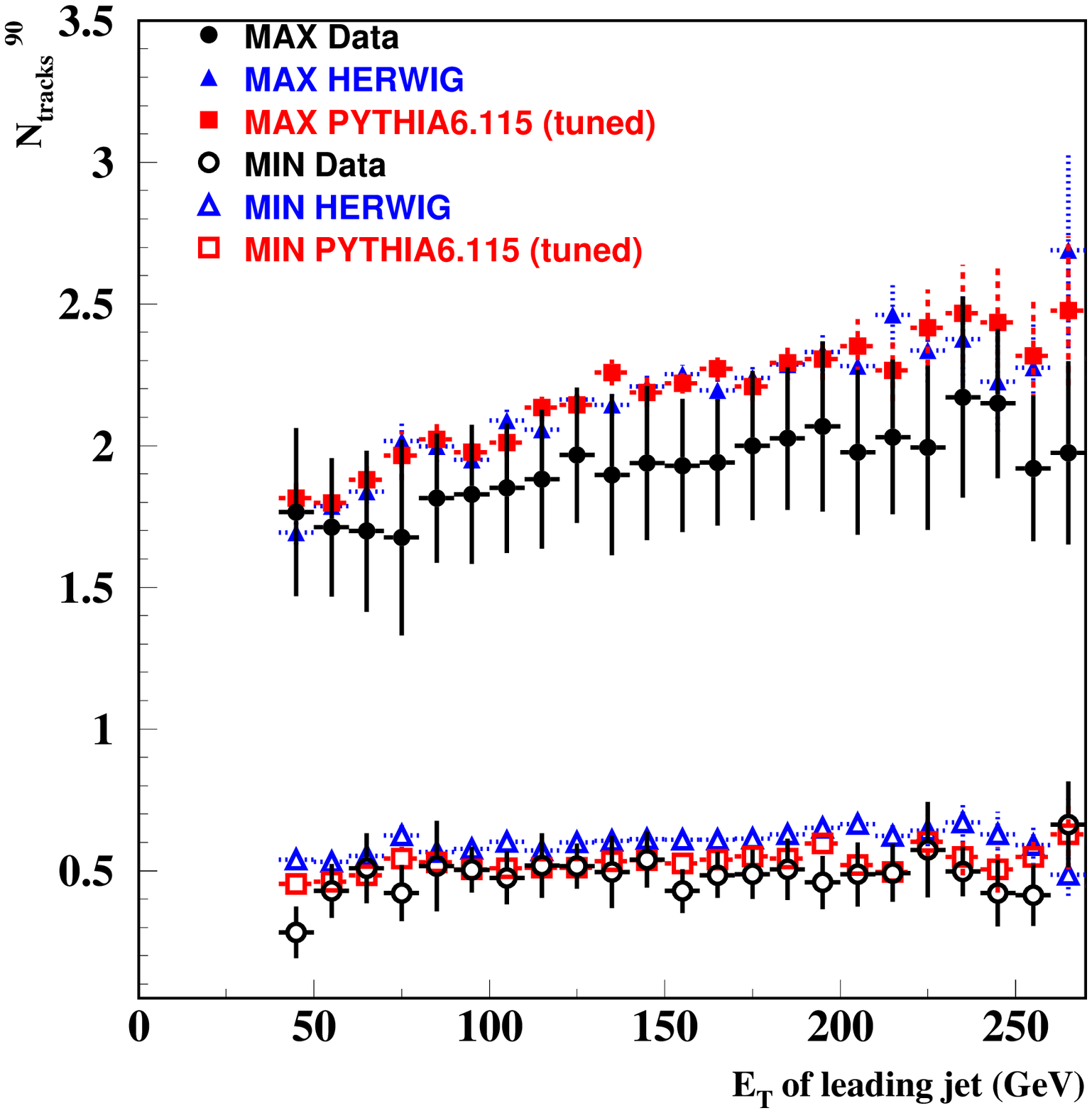}}
\caption{Number of tracks in the $max$ and $min$ cone as a function of the $E_T$ of the 
leading jet. Data, HERWIG and PYTHIA distributions are plotted at $\sqrt{s}=1800$ GeV. }
\label{Fig_ntrack_max_min}
\end{figure}
The average number of tracks found inside the two cones is shown in 
Figure~\ref{Fig_ntrack_max_min}, plotted as a function of $E_T^{(1)}$. 
A slightly higher track multiplicity is observed in both of the 
simulations compared to the data.

In Table~\ref{table_mb_pt}, the mean values  of the total track $P_T$ and
the mean number of tracks inside a cone randomly placed in the region
$|\eta|\le 0.5$ are shown for all minimum bias events and for those
with a high quality vertex only.
For the entire sample, the mean transverse momentum 
($P_T^{MB,cone}$) in the cone  is
about $0.36\pm 0.04$ GeV/c, 
while restricting the sample to events having a high quality vertex, the transverse momentum increases to $0.57\pm 0.06$ GeV/c. 
The average for the $P_T^{90,min}$ cone over the measured $E_T^{(1)}$ range is 
approximately 0.45 GeV/c, or midway between the above values.
\begin{table}[hb]
\begin{center}
\caption{Mean $P_T^{MB,cone}$  and the mean number of tracks in a random cone of radius 0.7 in
$\sqrt{s}=1800$ GeV minimum bias data. Only systematic errors are shown. 
Statistical errors are less than 0.5\%. }
\label{table_mb_pt}
\begin{tabular}{|c|c|c|c|}
\hline
\hline
\multicolumn{2}{|c|}{ } &
\multicolumn{1}{|c|}{$P_T^{MB,cone}$} &
\multicolumn{1}{|c|}{Track } \\
\multicolumn{2}{|c|}{ } &
\multicolumn{1}{|c|}{(GeV/c)} &
\multicolumn{1}{|c|}{Multiplicity} \\
\hline
\hline
DATA     & all vertices       & 0.36 $\pm$ 0.04 & 0.45 $\pm$ 0.06  \\
         & high quality vertex& 0.57 $\pm$ 0.06 & 0.69 $\pm$ 0.09   \\\hline
\multicolumn{2}{|l|}{HERWIG}   & 0.31            & 0.44             \\\hline
\multicolumn{2}{|l|}{PYTHIA (tuned)} & 0.35      & 0.44             \\\hline
\end{tabular}

\end{center}
\end{table}
\begin{figure}[h]
\centering
\mbox{
\hspace{-1.2cm}
\includegraphics[height=8.5cm]{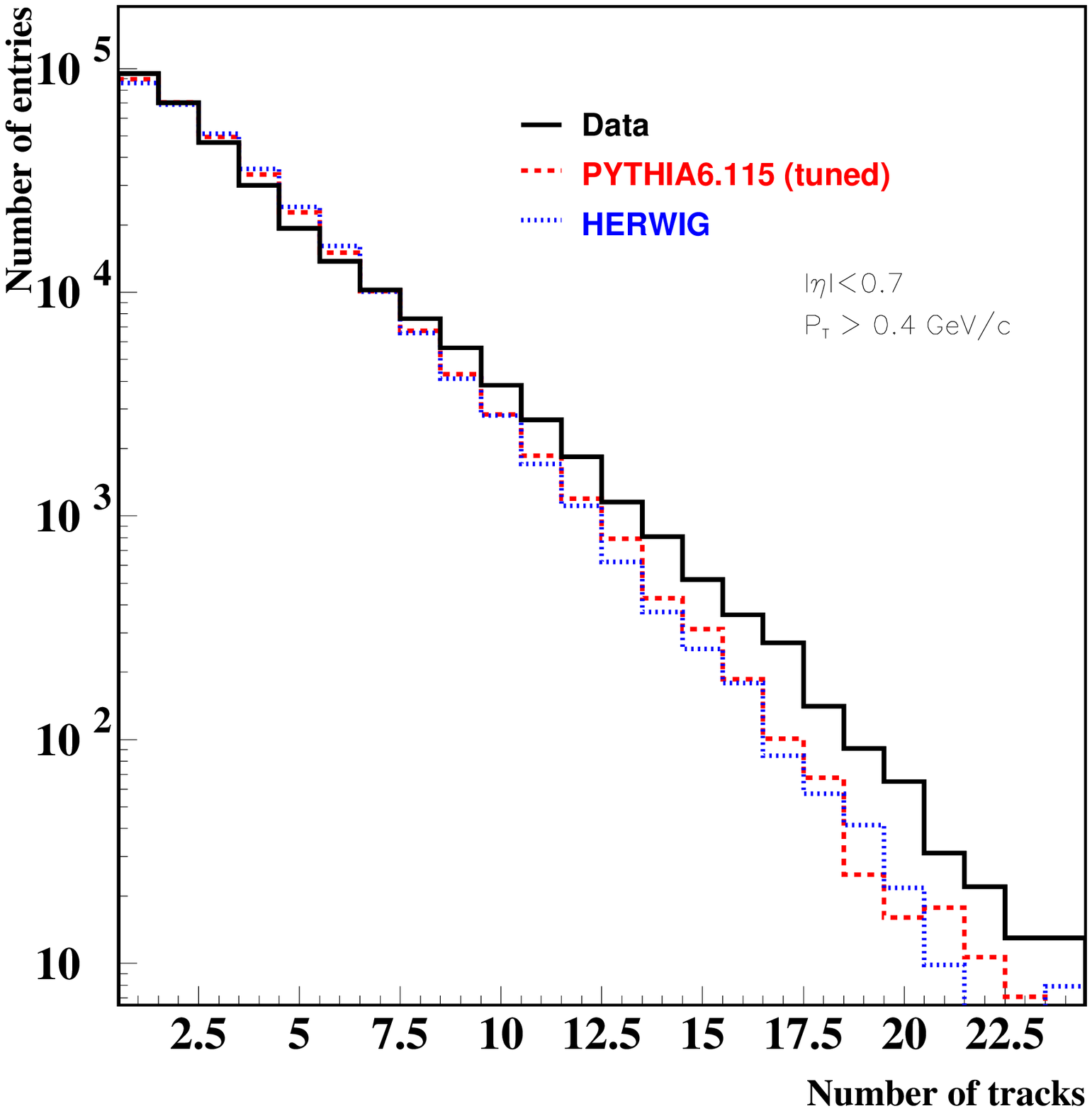}
\includegraphics[height=8.5cm]{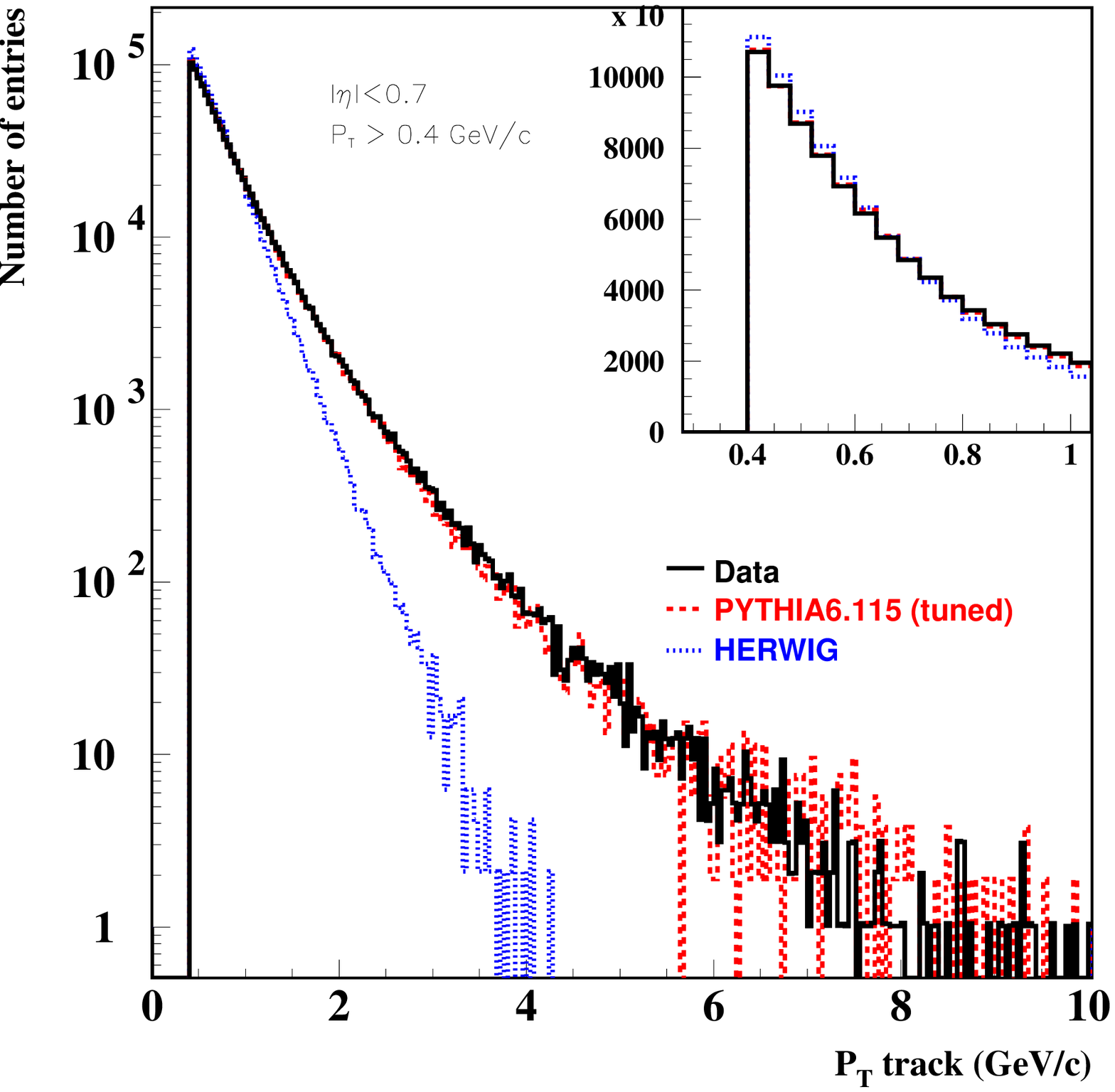}
}
\caption{Distribution of track multiplicity (left) and  transverse momentum (right) in the $\sqrt{s}=1800$ GeV minimum bias sample. The inset in the right plot shows the low $P_T$ portion of the spectrum on a linear scale.} 
\label{Fig_pt_her}
\end{figure}

The track multiplicity and track momentum distributions in minimum bias data 
are shown in 
Figure~\ref{Fig_pt_her}, with the number of entries in the simulation normalized to 
the  number in the data. The transverse momentum distribution at high $P_T$ is not well-reproduced
by HERWIG, which has virtually no tracks with $P_T\ge  4$ GeV/c. 
The absence of high $P_T$ tracks indicates the lack of a semi-hard
processes in the HERWIG model of minimum bias events. In contrast, PYTHIA
reproduces the transverse momentum distribution considerably better. The model
of multiple parton interactions incorporated in the PYTHIA description of
minimum bias events~\cite{dps} allows for the possibility of high transverse momentum
tracks. Neither HERWIG nor PYTHIA appears to correctly describe the  high
multiplicity end of the track multiplicity distribution.

As  previously described, the {\it Swiss cheese} distribution is formed by summing the
transverse momentum of the tracks in the central region ($|\eta|\le 1$), excluding the
transverse momentum of the tracks in a radius 0.7 from the center of the two (or
three) most energetic jets in the event (where an $E_T$ requirement of 5 GeV has been placed on each jet).

\begin{figure}[h]
\centerline{\includegraphics[height=11cm]{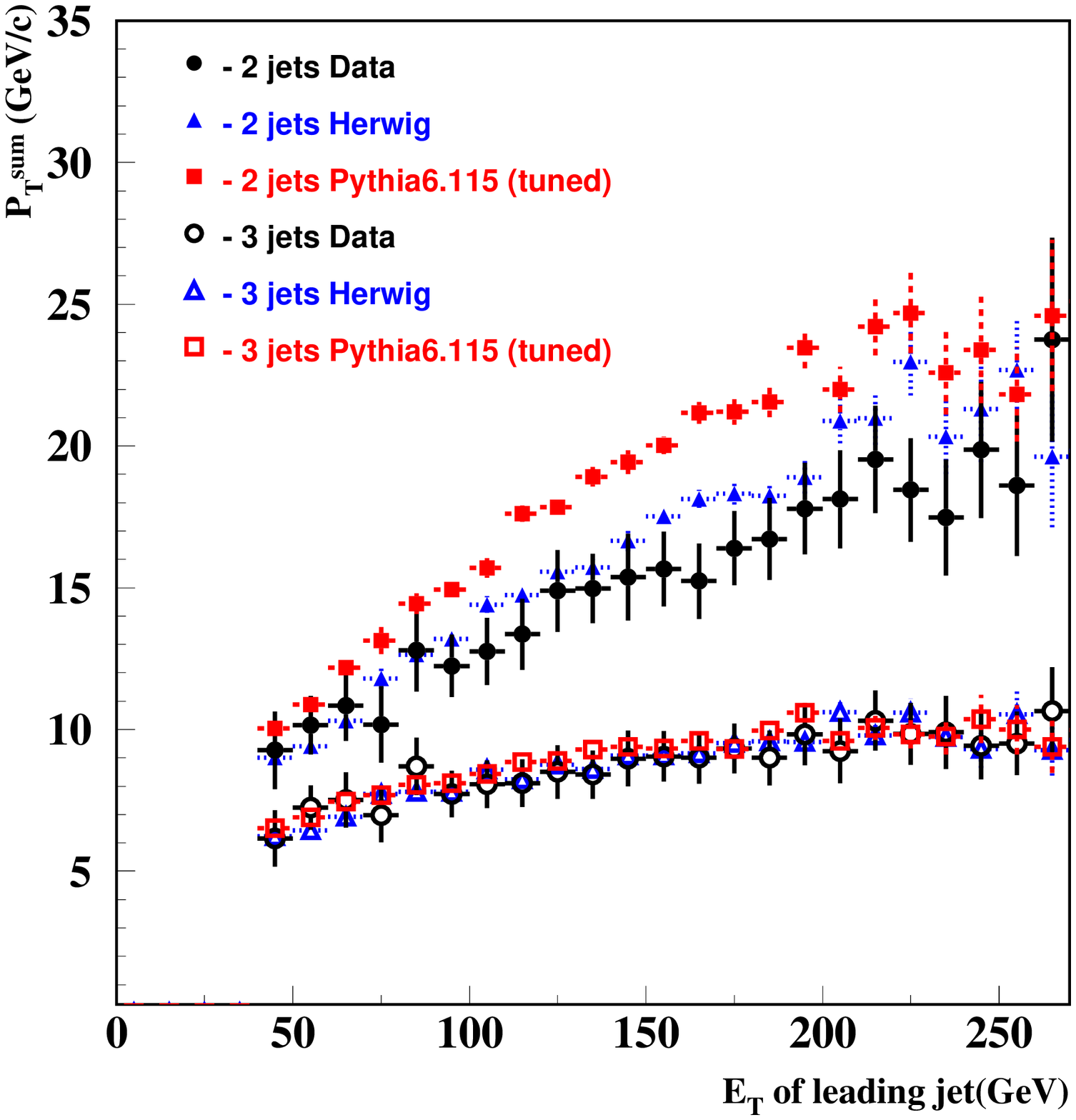}}
\caption{$P_{T}^{sum}$ ({\it Swiss cheese}). The two and three most energetic jets in each event 
are subtracted from the total transverse momentum in the central detector region. 
Data, HERWIG and PYTHIA results are shown at $\sqrt{s}=1800$ GeV.}
\label{Fig_swiss_cheese_pt}
\end{figure}

The sum of the track transverse momentum in the 
central region for the 2-jet subtracted and 3-jet subtracted distributions is shown in Figure~\ref{Fig_swiss_cheese_pt} for the data,  HERWIG and the version of PYTHIA tuned for best agreement with the $max$/$min$ cone data. In the simple picture presented earlier,  the difference between the {\it Swiss cheese} level with the 2 highest $E_T$ jets subtracted and the corresponding minimum bias level should be proportional to the NLO  (third parton) and higher order contributions. The {\it Swiss cheese} level with the three highest  $E_T$ jets subtracted should have little or no NLO contribution. 
 Both the 2-jet subtracted and 3-jet subtracted distributions increase as 
the lead jet $E_T$  increases,  with the slope being less for the 3-jet subtracted  case.    The 3-jet subtracted {\it Swiss cheese} average $P_T$/(unit $\eta$--$\phi$) is 
$0.92\pm0.09 $ GeV/c  compared to $0.37\pm 0.04$ GeV/c observed in  minimum bias data with high quality vertices and $0.23\pm 0.04$ GeV/c in all minimum bias events. The larger momentum observed in the 3-jet subtracted  {\it Swiss cheese} distribution indicates additional contributions than just those  from the soft underlying event. These contributions include hadronization from the jets ({\it splash-out}), double-parton scattering, higher order  radiation effects~\cite{pumplin}, as well as contributions from $3^{rd}$ jets that fail the $E_T$ threshold cut of 5 GeV. For comparison, the average
momentum/(unit $\eta$--$\phi$) in the $min$($max$) cone is  $0.29\pm 0.04$ ($1.91\pm 0.14$) GeV/c. 

\begin{figure}[h]
\centering
\mbox{
\includegraphics[height=11.cm]{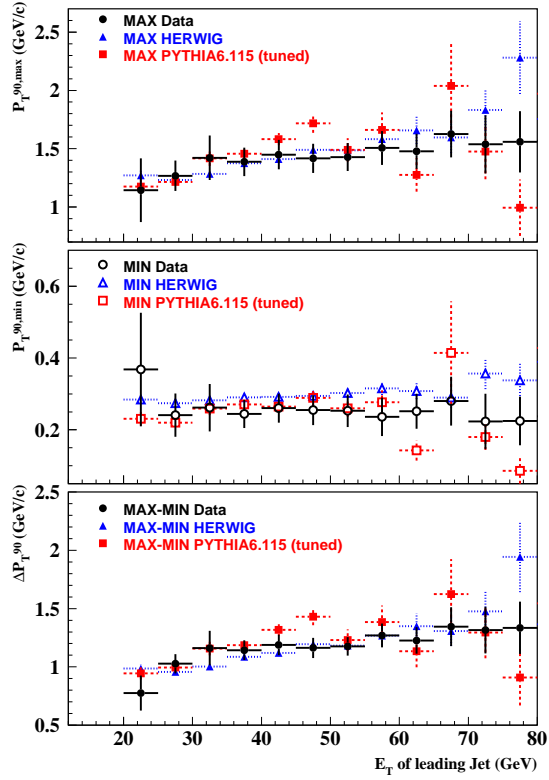}}
\caption{$P_T^{90,min}$, $P_T^{90,max}$ and their difference $\Delta P_T^{90}$  at $\sqrt{s}= 630$ GeV as a function of the $E_T$ of the leading jet.}
\label{Fig_track_max_min_630} 
\end{figure}

To study the energy dependence, we have analyzed jet and  minimum bias 
data at $\sqrt{s}=630$ GeV. 
In Figure~\ref{Fig_track_max_min_630}, $P_T^{90,max}$ and  $P_T^{90,min}$ 
(and their difference) are plotted as a function of $E_T^{(1)}$.
The $\sqrt{s}=630$ GeV data shows a similar behavior as was observed at $\sqrt{s}=1800$ GeV but the overall
magnitudes are lower. The average $P_T^{90,min}$ ($P_T^{90,max}$) at $\sqrt{s}=630$ GeV is  
$ 0.25\pm0.04 $ ($ 1.43\pm0.12 $) GeV/c, 
$\sim 0.2$ ($\sim 1.5$)GeV/c lower than what is observed at $\sqrt{s}=1800$ GeV. 
Both HERWIG and PYTHIA reproduce the data at 630 GeV well.
PYTHIA has been tuned as for the analysis at 1800 GeV, but with the regularization
scale, $P_{T0}$, set to 1.4 GeV/c. 
A dependence of $P_{T0}$ on the center of mass energy has been
implemented in versions of PYTHIA after 6.12  according to the model
described in~\cite{pythia6.12}. This  model, however,  predicts a value for $P_{T0}$ of 1.9 GeV/c for the 630
GeV data while in the 1800 GeV data the prediction is  2.3 GeV/c, showing a smaller
dependence on the center-of-mass energy than that observed in our data.
Using the default values the PYTHIA predictions underestimates the
number of charged particles at $\sqrt{s}=630$ GeV.

\begin{figure}[h]
\centerline{\includegraphics[height=11.cm]{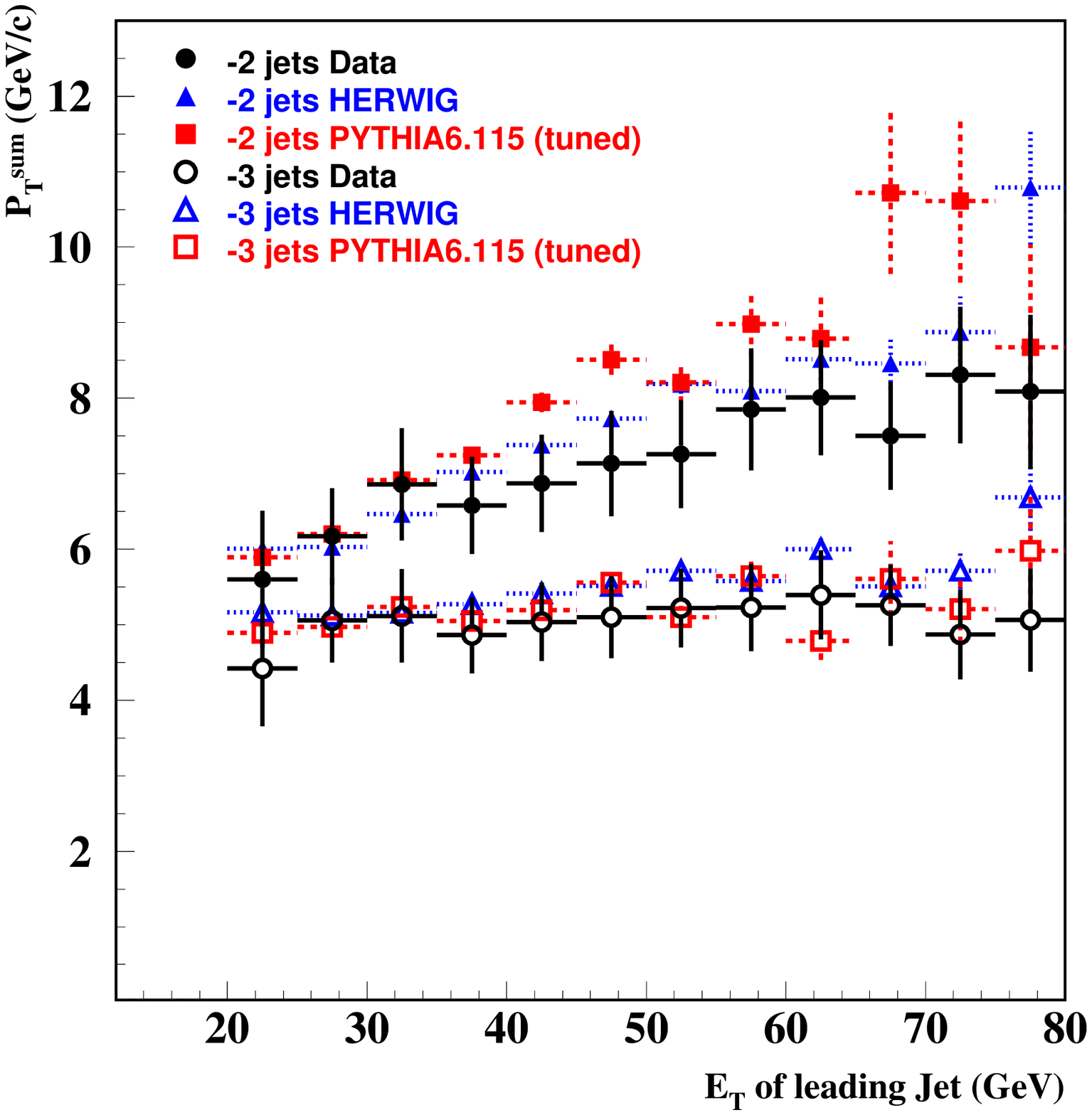}}
\caption{$P_{t}^{sum}$ ({\it Swiss cheese}). The two and three most energetic jets in the
events are subtracted from the total transverse momentum in the central detector 
region. Data, HERWIG and PYTHIA results are shown  at $\sqrt{s}= 630$ GeV. }
\label{Fig_track_swisscheese_630}
\end{figure}

The {\it Swiss cheese} distributions at 630 GeV are shown in Figure~\ref{Fig_track_swisscheese_630}. A very good
agreement between data and both Monte Carlos is observed if the three most
energetic jets are subtracted. In the case where the two most
energetic jets are subtracted, both Monte Carlos lie above the data, as was also 
observed at 1800 GeV. Again, the momentum is larger when three jets are subtracted  
($P_T$/(unit $\eta$--$\phi$) is $0.52\pm $0.05 GeV/c ) than in  minimum bias events with a high quality vertex($P_T$/(unit $\eta$--$\phi$) is $0.34\pm $0.03 MeV/c).

\begin{figure}[ht]
\centering
\mbox{
\includegraphics[height=8.5cm]{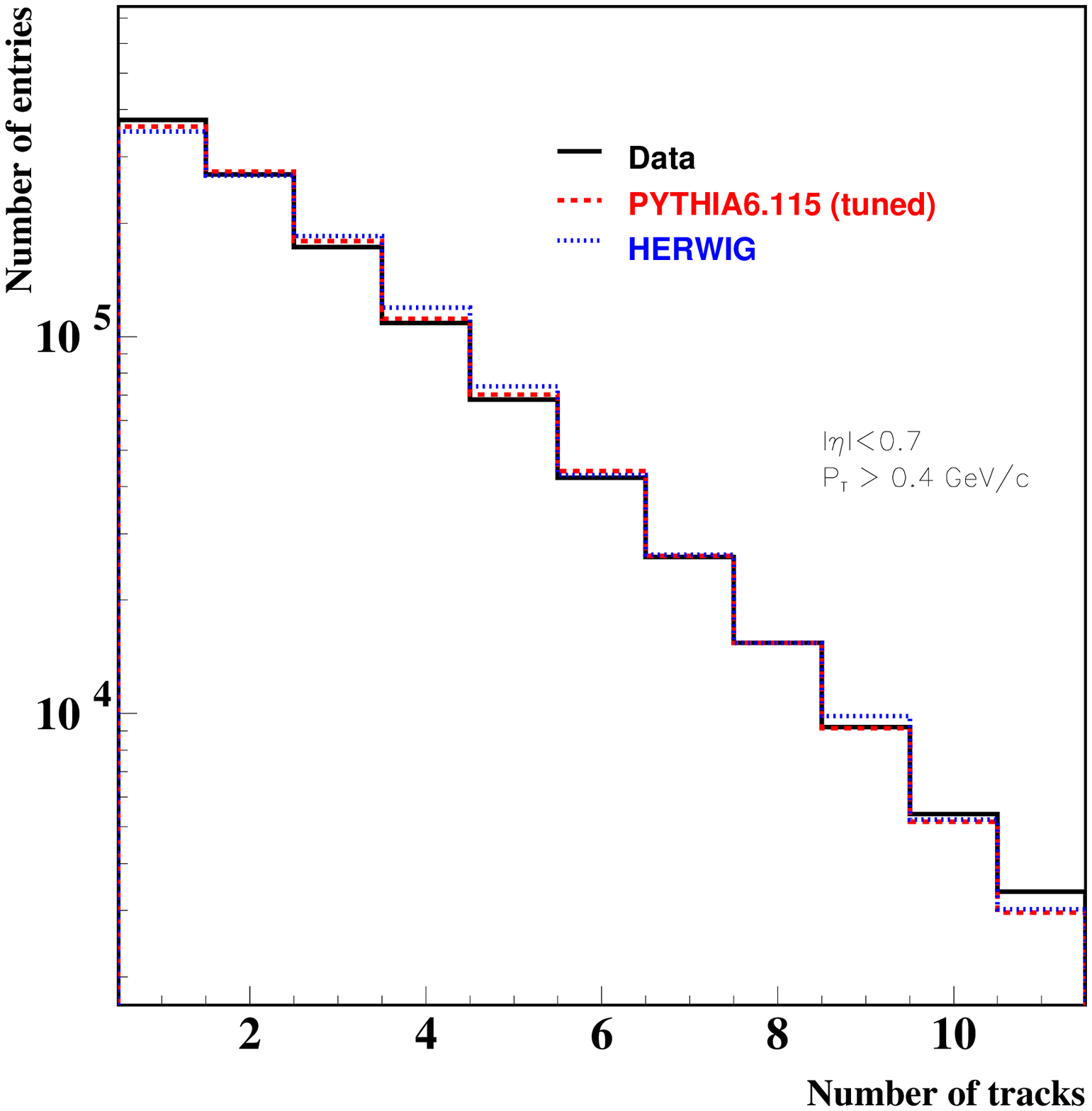}
\includegraphics[height=8.5cm]{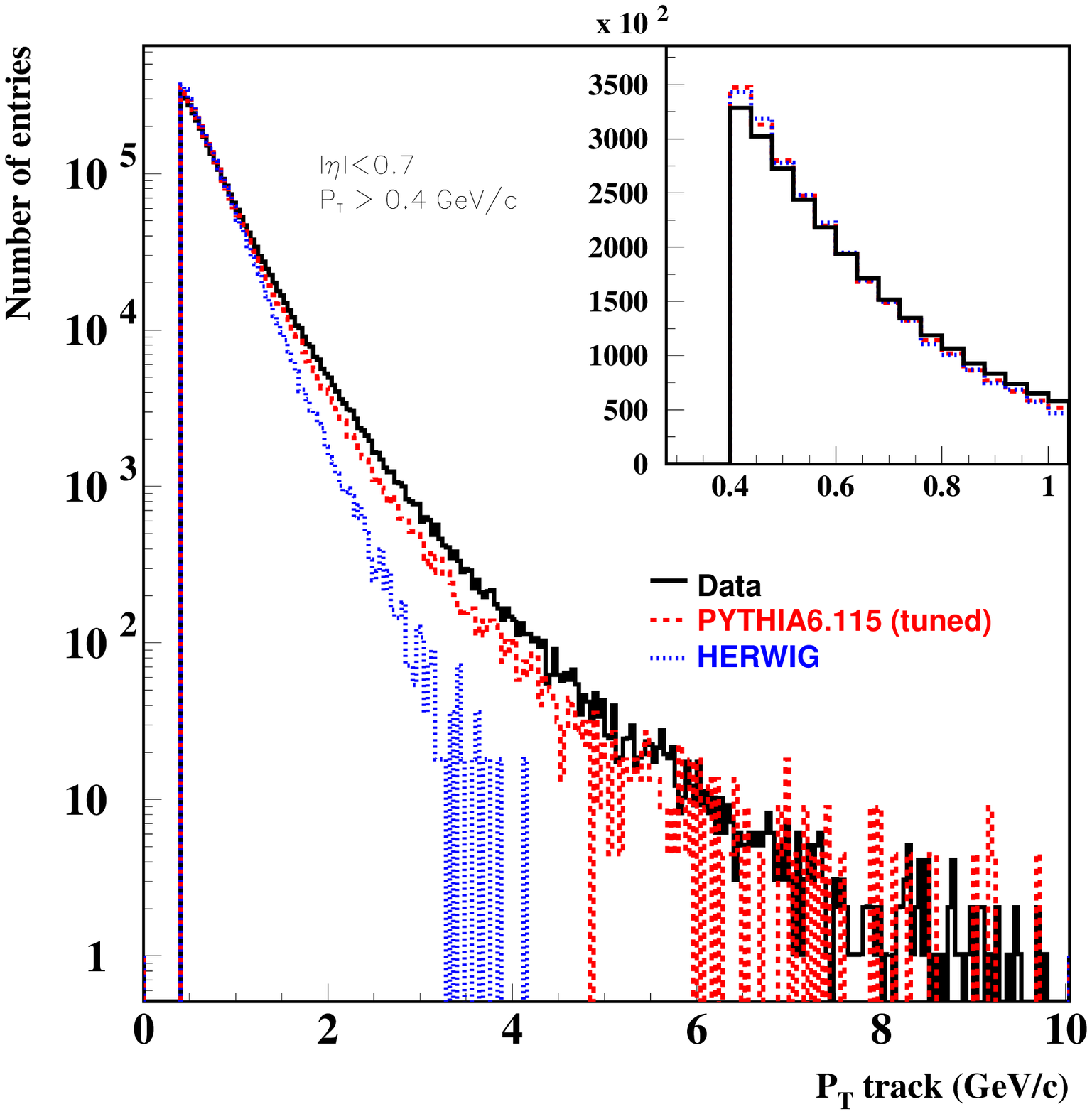}
}
\caption{Distribution of track multiplicity (left) and  transverse momentum (right) in the $\sqrt{s}=630$ GeV minimum bias sample. The inset in the right plot shows the low $P_T$ portion of the spectrum on a linear scale.} 
\label{Fig_pt_her_630}
\end{figure}

In Table~\ref{Tab_mb_max_min_track_630} are shown the average value of the  total track $P_T$ and the mean number of tracks inside
a cone in the central rapidity region in minimum bias data at $\sqrt{s}=630$ GeV.  
The sum of the track transverse momenta  is 
$20\%$ lower with respect to the 1800 GeV sample. Figure ~\ref{Fig_pt_her_630} shows the 
track multiplicity  and momentum distributions for minimum bias
events. Again, the number of entries in the simulation is
normalized to the number of entries in the data.
The track multiplicity distribution and the mean $P_T^{MB,cone}$, dominated by the low edge of the steeply falling
spectrum, is well reproduced by both Monte Carlo generators. 
Unlike the situation at 1800 GeV, PYTHIA fails to produce enough high $P_T$ tracks, although it 
still produces considerably more than HERWIG.

\begin{table}[!hb]
\begin{center}
\caption{Data and simulation comparisons for minimum bias events at $\sqrt{s} = 630$ GeV. 
Average $P_T^{MB,cone}$  and the average number of tracks in a random cone of radius 0.7 are shown. Only systematic errors are shown. Statistical errors are less than 0.2\%. }
\begin{tabular}{|c|c|c|c|}
\hline
\hline
\multicolumn{2}{|c|}{} &
\multicolumn{1}{|c|}{$P_T^{MB,cone}$} &
\multicolumn{1}{|c|}{Track} \\
\multicolumn{2}{|c|}{} &
\multicolumn{1}{|c|}{(GeV/c)} &
\multicolumn{1}{|c|}{Multiplicity} \\
\hline
\hline
DATA     & all vertices &0.29 $\pm$ 0.03 & 0.37 $\pm$ 0.05 \\
         & high quality vertex & 0.52 $\pm$ 0.05 & 0.65 $\pm$ 0.08\\\hline
\multicolumn{2}{|l|}{HERWIG}   & 0.26 & 0.36 \\\hline
\multicolumn{2}{|l|}{PYTHIA}   & 0.28 & 0.38 \\\hline
\end{tabular}
\label{Tab_mb_max_min_track_630}
\end{center}
\end{table}

In summary we have studied the momentum deposited in two cones at $\pm 90^\circ$ to the 
highest $E_T$ jet in hard interaction events at $\sqrt{s}=1800$ and 630 GeV. The maximum of the 
two cone energies increases with highest $E_T$ jet in the event whereas  the minimum is
flat. Both HERWIG and PYTHIA exhibit the same behavior but HERWIG provides a better description of the CDF data. The
momentum in the $min$ cone is
midway between the levels observed in generic minimum bias events and minimum bias events selected by high track multiplicity. 
In the HERWIG minimum bias model, the generated tracks are too soft, and semi-hard or hard interactions 
should be added to the minimum bias events in order to better reproduce the data.
PYTHIA, however, with an adequate tuning of its parameters, 
reproduces the charged particle distribution better than HERWIG for the 1800 GeV minimum bias data,  but is less successful at 630 GeV. 
We have measured the $\sqrt{s}$ dependence of the underlying event momentum in jet events and ambient
energy in minimum bias events. The underlying momentum at $\sqrt{s}=630$ GeV is $45\%$ lower than 
what is observed  in 1800 GeV data.
These measurements will allow for more precise tunings of both the underlying event in Monte Carlo programs and the mechanisms for gluon radiation, help to  reduce the uncertainties in future jet studies at the Tevatron
and will lead to a better prediction of physics signals and backgrounds at the Large Hadron Collider. 

\section{Acknowledgement}

We thank the Fermilab staff and the technical staffs of the participating 
institutions for their vital contributions. This work was supported by the 
U.S. Department of Energy and National Science Foundation; 
the Italian Istituto Nazionale di Fisica Nucleare; 
the Ministry of Education, Culture, Sports, Science and Technology of Japan; 
the Natural Sciences and Engineering Research Council of Canada; 
the National Science Council of the Republic of China; 
the Swiss National Science Foundation; 
the A.P. Sloan Foundation; 
the Bundesministerium fuer Bildung und Forschung, Germany; 
the Korean Science and Engineering Foundation and 
the Korean Research Foundation; 
the Particle Physics and Astronomy Research Council and the Royal Society, UK; 
the Russian Foundation for Basic Research; 
the Comision Interministerial de Ciencia y Tecnologia, Spain; 
in part by the European Community's Human Potential Programme under contract 
HPRN-CT-20002, Probe for New Physics; 
and by the Research Fund of Istanbul University Project No. 1755/21122001.



\end{document}